\documentclass[reprint,amsmath,amssymb,aps]{revtex4-2}
\usepackage{graphicx}
\usepackage{subfigure}
\usepackage{dcolumn}
\usepackage{bm}
\usepackage{hyperref}
\usepackage{indentfirst}
\usepackage{braket}
\usepackage{amsmath}
\usepackage{color}
\usepackage{makecell}
\usepackage{booktabs}
\usepackage{footnote}
\usepackage{longtable}
\usepackage[ruled,vlined]{algorithm2e}
\hypersetup{colorlinks=true, citecolor=blue, urlcolor=blue, linkcolor=blue}

\begin{document}
\title{Weak Fragmentation and Thermalization in a Dipole-Conserving Bose-Hubbard Chain}
\author{Chenrong Liu}
\email{crliu14@fudan.edu.cn}
\affiliation{Department of Physics, Yichun University, Jiangxi 336000, China}
\affiliation{College of Mathematics and Physics, Wenzhou University, Zhejiang 325035, China}
\affiliation{State Key Laboratory of Surface Physics and Department of Physics, Fudan University, Shanghai 200433, China}

\begin{abstract}
We study Hilbert-space fragmentation and thermalization in a one-dimensional dipole-conserving Bose-Hubbard chain. By analyzing the structure of the Hamiltonian matrix in the Fock basis, we show that the system exhibits weak Hilbert-space fragmentation. We further construct an exponentially large family of frozen product states and derive analytical upper and lower bounds on their number. Using exact diagonalization, we examine the consequences of weak fragmentation for eigenstate half-chain entanglement, density relaxation dynamics, and level statistics. All these quantities reveal a transition from a weak eigenstate thermalization regime to a nonergodic regime with increasing on-site interaction strength. These results show that weak Hilbert-space fragmentation \textit{does not} preclude quantum chaos or thermalization, and provides a minimal platform for studying the interplay of dipole conservation, weak fragmentation, and ergodicity breaking.
\end{abstract}

\maketitle

\section{Introduction} 

Understanding when quantum many-body systems thermalize has become a central problem in nonequilibrium physics. In interacting systems, thermalization is commonly explained by the eigenstate thermalization hypothesis (ETH), which demonstrates that many-body eigenstates already encode thermal expectation values of local observables \cite{PhysRevA.43.2046,PhysRevE.50.888,Nature7189,PhysRevE.60.3949}. When ETH holds, an initially out-of-equilibrium state loses memory of microscopic details under unitary evolution, and long-time observables are determined only by global conservation laws. By contrast, violations of ETH can arise from integrability, disorder induced many-body localization, quantum scars, or strong dynamical constraints \cite{Essler_2016, MBL1,MBL2,MBL3,Scar1,Scar2,Scar3,Scar4,Scar5,PhysRevLett.124.207602,PhysRevResearch.2.023159}. Identifying new mechanisms for ETH holding and violation in interacting systems remains an active area of research. 

A particularly important class of constrained dynamics is the systems that have multipole conservation laws. In recent years, dipole-conserving models have attracted broad interest in connection with fracton-like dynamics, constrained transport, and slow relaxation \cite{PhysRevLett.132.220405,PhysRevE.107.034142,PhysRevB.101.214205,dipoleFermion,PhysRevLett.121.040603}. In such systems, the conservation of particle number and dipole moment strongly limits the allowed local hopping processes. As a result, a single particle can not move freely, and transport can proceed only through the dipole hopping that preserves the center of mass. This constraint can produce many unusual many-body phenomena, including fractonic phases, special dynamical spectral response, fractonic fluctuations, and deconfinement dynamics of fractons\cite{PhysRevB.109.125137, PhysRevLett.132.143401,PhysRevB.106.064511,PhysRevB.107.195131,PhysRevB.107.195132,PhysRevLett.132.143401,PhysRevResearch.6.023269,vsy9-f5yn}.

One of the most striking consequences of dipole conservation is Hilbert-space fragmentation \cite{HSF3,PhysRevLett.130.010201,PhysRevX.12.011050,SciPostPhys.18.3.111,PhysRevB.109.064302,HSF2,PhysRevB.108.045127,PhysRevX.10.011047}. In a fragmented system, even after resolving all global symmetries, the Hilbert space can split into many dynamically disconnected Krylov sectors that cannot be connected by repeated action of the Hamiltonian. This is a stronger notion than ordinary symmetry resolution, because it originates from local dynamical constraints rather than from a conventional conserved quantity generated by a global operator. Depending on the model, the fragmentation may be strong, with many comparable sectors, or weak, with one giant connected component coexisting with many small disconnected ones \cite{PhysRevX.10.011047}. This distinction is important because weakly fragmented systems may still display thermal behavior for most states in the dominant sector, even though nonthermal eigenstates remain present.

However, most previous studies of dipole-conserving fragmentation have focused on spin models or related constrained lattice systems \cite{PhysRevX.10.011047,SciPostPhys.14.6.140,SciPostPhys.13.4.098}.  While bosonic realizations with exact dipole conservation are less explored \cite{Nature0.1038} and raise several natural questions. Does a minimal dipole-conserving Bose-Hubbard chain exhibit strong or weak fragmentation? How abundant are frozen states in the physically symmetric sectors? Can weak fragmentation coexist with a quantum chaotic energy spectrum and thermalization? Or can the dipole constraint prevent ergodicity? Addressing these questions is also helpful for clarifying which features are universal results of dipole conservation and which depend on the microscopic realization of the constrained dynamics.

In this work, we study a one-dimensional dipole-conserving Bose-Hubbard model. Focusing on the symmetry sector with fixed total particle number and zero total dipole moment, we first analyze the Hamiltonian matrix structure in the many-body Fock basis. We find that the Hilbert space fragments into exponentially many Krylov sectors, but the fragmentation is weak. We further identify exponentially many frozen product states and derive analytical upper and lower bounds. Then, we investigate the implications of this weak fragmentation for thermalization and quantum chaos using exact diagonalization. By calculating several quantities, we show that the system satisfies the weak ETH on the accessible system sizes for large dipole hopping amplitudes, and it displays nonergodic behavior for the otherside. Our results therefore show that the dipole-conserving bosonic model provides a simple example in which weak Hilbert-space fragmentation coexists with the quantum chaotic and nonergodic regimes.

This paper is organized as follows. In Sec .~\ref {Model}, we introduce the dipole-conserving Bose-Hubbard chain and specify the symmetry sector studied numerically. In Sec .~\ref {fragmentation}, we analyze the fragmented structure of the Hilbert space and the counting of frozen states. In Sec .~\ref {ETH} and Sec .~\ref {Eigval}, we examine entanglement entropy, real-time autocorrelations, and level statistics to characterize the evolution from weak ETH behavior to nonergodicity. Technical details of the frozen state counting and related numerical procedures are presented in the Appendices.

\section{Dipole-conserving Bose-Hubbard chain} \label{Model}
\begin{figure}[htbp]
\begin{center}
\includegraphics[width=0.45\textwidth]{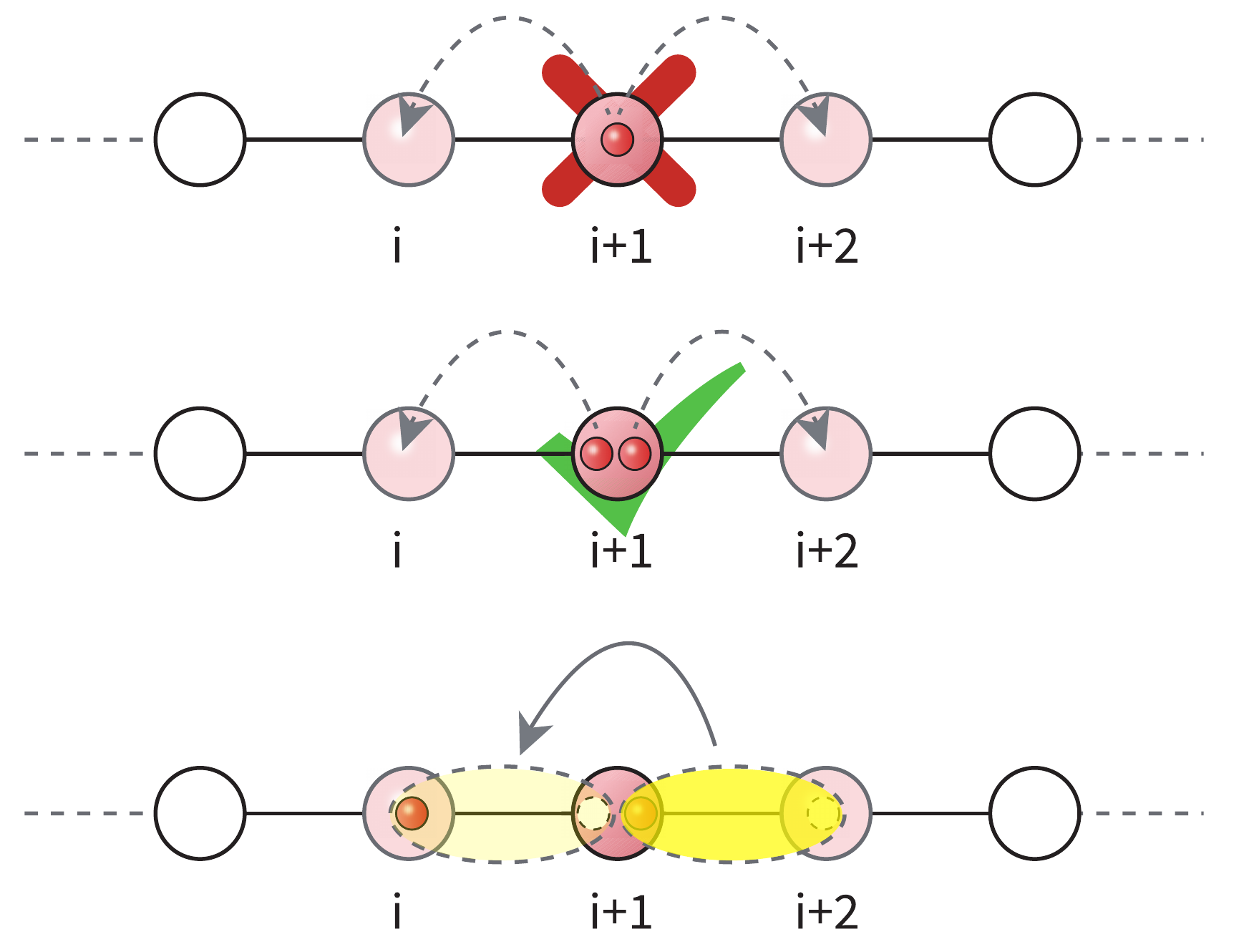}
\caption{Illustration of the restricted hopping process $\hat{b}_i^{\dagger} \hat{b}_{i+1}^2 \hat{b}_{i+2}^{\dagger}$ which conserves dipole moment. The standard single-particle hopping is forbidden (top), but two particles simultaneously move from the site $i+1$ to nearest neighbor sites $i$ and $i+2$ is allowed (middle). This mechanism is equivalent to the free propagation of a dipole, visualized here as a bound boson-hole pair denoted by a shadow ellipse (bottom). }
\label{Fig1}
\end{center}
\end{figure}

Let us first introduce the model in this part.  We study a system of interacting bosons on an open boundary chain with length $L$. The system is described by the Hamiltonian $\hat{H}$, which conserves both the total particle number $\hat{N}=\sum_i\hat{n}_i$, and the total dipole moment (or center of mass) $\hat{P}=\sum_ii\hat{n}_i$. The Hamiltonian is defined as,
\begin{equation} \label{H}
\hat{H}=- \sum_i J_i\left(\hat{b}_i^{\dagger} \hat{b}_{i+1}^2 \hat{b}_{i+2}^{\dagger}+\text { h.c. }\right)+\frac{U}{2} \sum_i \hat{n}_i\left(\hat{n}_i-1\right),
\end{equation}
where $J_i >0$ is the two-particle hopping amplitude, $U>0$ is the on-site repulsive interaction strength,  $\hat{b}_i$ ($\hat{b}_i^\dag$) is the boson annihilation (creation) operator on site $i$, and $\hat{n}_i=\hat{b}^\dag_i\hat{b}_i$ is the local number operator. A sketch of the hopping process of Hamiltonian  (\ref{H}) has been placed in Fig.~\ref{Fig1}. 

In contrast to the standard Bose-Hubbard model, single-particle hopping is forbidden in this system due to the conservation of the dipole moment. The kinetic term $\hat{b}_i^{\dagger} \hat{b}_{i+1}^2 \hat{b}_{i+2}^{\dagger}$ represents a correlated hopping process in which two particles simultaneously tunnel from site $i+1$ to the nearest-neighbor sites $i$ and $i+2$, a mechanism that ensures dipole conservation. For this reason, a single particle can move only through the emission or absorption of a dipole.  Furthermore, the Hamiltonian (\ref{H}) is invariant under the local gauge transformation:
\begin{equation}
\hat{b}_i^\dag \rightarrow e^{-j \frac{\pi}{2}i^2}\hat{b}_i^\dag,
\end{equation}
where $j$ is the imaginary unit. This implies that $\hat{H}(J_i)$ and $\hat{H}(-J_i)$ have the same energy spectrum. 

This model can also be understood as a system of interacting dipoles. By defining the dipole operator as $\hat{d}_i=\hat{b}_i^\dag\hat{b}_{i+1}$, the hopping term of the Hamiltonian can be formally rewritten to highlight the dipole dynamics. We note that the operators $\hat{d}_i$ and $\hat{d}_i^\dag$ do not obey the standard bosonic commutation relations, reflecting the constrained nature of the dynamics.

For numerical calculations, we take the length of the chain $L=2k+1$ ($k\in \mathbb{Z}^+$ ) to be odd, and label the sites symmetrically as $i=-k,\cdots,0,\cdots,k$, centered at site $0$ (the reference site). Strictly speaking, the sum in the first term of Hamiltonian (\ref{H}) is $-\sum_{i=-k}^{k-2}J_i(\cdots)$.  We choose the hopping amplitudes to be fixed at
\begin{equation}
J_i=\left\{\begin{array}{cl}
2J, & i =-k \\
J, & i>-k \end{array}\right.
\end{equation}
where $J=1$ is set to be the energy unit in our calculations. This choice can let us explicitly break the spatial reflection symmetry of the chain, ensuring that the resulting energy spectrum belongs to a single symmetry class.  

In the presented paper, we focus on the symmetry sector with the total particle number fixed at $N=2L$ (a filling of two bosons per site) and the total dipole moment fixed at $P=0$. For the numerical analysis of the Hilbert-space fragmentation, we can use $L$ up to 13. While for the analysis of other quantities, we use the full matrix exact diagonalization (ED) scheme to obtain all eigenvalues and eigenstates for different values of $U$. Due to the computation limit, the chain length $L$ is up to 9, and the matrix dimension is up to 33885 for full ED.

\section{Weak Hilbert space fragmentation} \label{fragmentation}
In this part, we will show that the dipole-conserving hopping term can lead to the Hilbert space fragmentation, which means that within a given symmetry sector, the Hilbert space fragments into many (often exponentially many) dynamically disconnected invariant subspaces (Krylov sectors) that are not connected by the Hamiltonian’s off-diagonal terms\cite{HSF3,PhysRevLett.130.010201,PhysRevX.12.011050,SciPostPhys.18.3.111,PhysRevB.109.064302,HSF2,PhysRevB.108.045127,PhysRevX.10.011047}. Here, the dipole-conserving constraint plays an important role, which severely restricts the set of allowed hops. Although the Hamiltonian preserves the global symmetries, dipole conservation further forbids many local hopping processes that would connect different Fock states. This makes the Hamiltonian matrix break into many disconnected components within a given symmetry sector. 

Previous work on the spin-1 version of Hamiltonian~(\ref{H}) showed that Hilbert-space fragmentation can be strong or weak, depending on whether the dipole-conserving spin interaction involves three-site or four-site processes \cite{PhysRevX.10.011047}. In particular, the minimal model with three-site spin terms exhibits strong fragmentation into exponentially many invariant subspaces in the local $S_z$ basis. In contrast, we demonstrate here that the bosonic realization with only three-site dipole-conserving hopping yields weak fragmentation in the symmetry sector $(N, P)=(2L,0)$. A weak fragmentation means that although the Hilbert space still decomposes into exponentially many Krylov sectors in the Fock basis, a single dominant block occupies an overwhelming fraction of the Hilbert space. 
\subsection{The Hamiltonian matrix structure}
\begin{figure}[htbp]
\begin{center}
\includegraphics[width=0.48\textwidth]{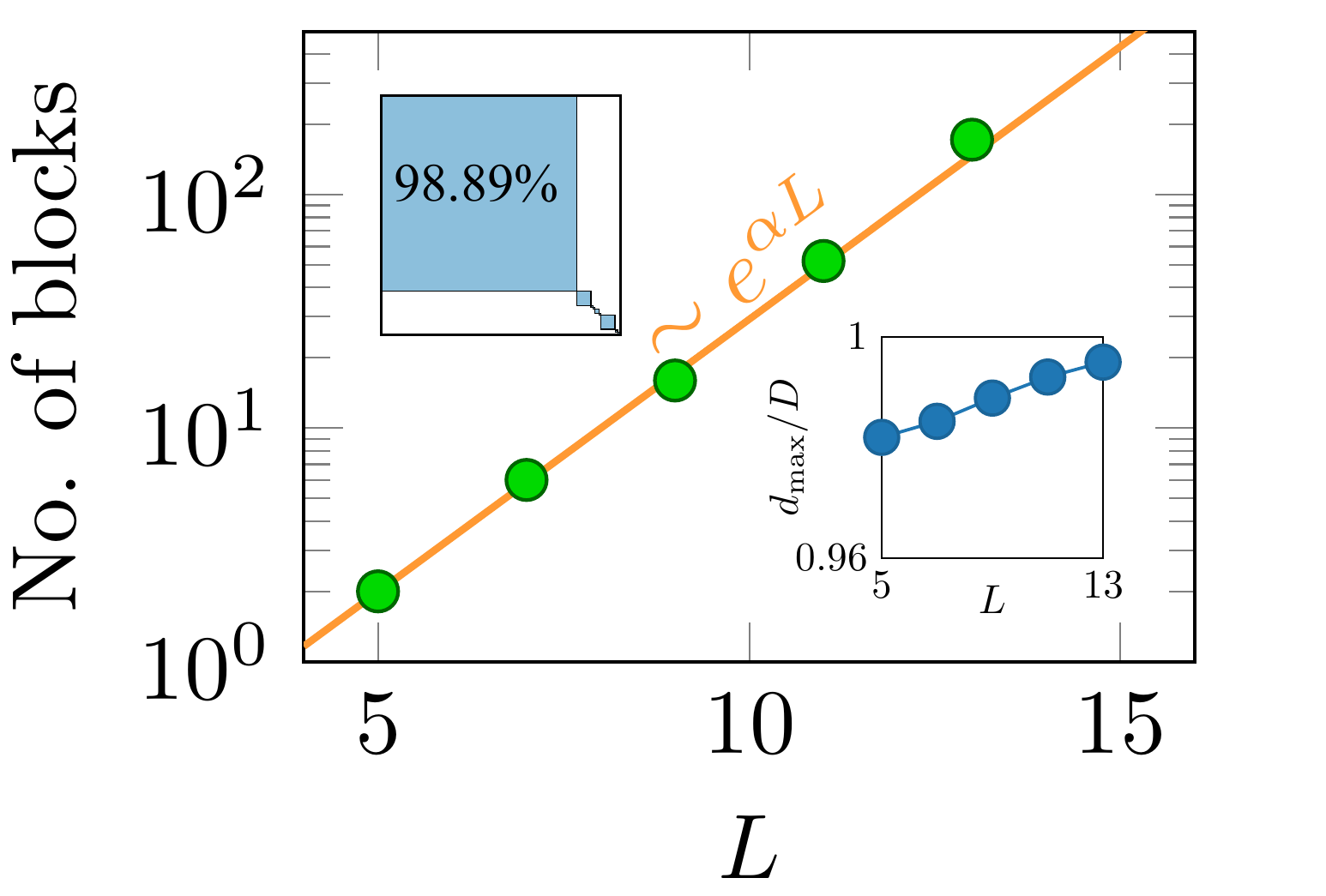}
\caption{(Color online) Hilbert-space fragmentation vs system size in the symmetry sector $(N, P)=(2L,0)$. We analyze fragmentation across lattice sizes $L$ up to 13, where the Hilbert-space dimension reaches $D=33427622$ and fragments into 172 disconnected blocks. The main panel shows that the number of disconnected blocks is exponentially growing with $L$. The orange curve is an exponential fit, $\sim\exp(\gamma L)$, yielding $\gamma\approx0.538$. Upper inset: an example Hamiltonian matrix in the Fock basis for $L=9$, illustrating the fragmentation structure; the largest block contains $98.89\%$ of the total dimension (its scale is rescaled for visual guidance). Lower inset: the largest block dimension $d_\text{max}$ scaled with the total Hilbert-space dimension $D$ as a function of $L$, suggesting it can approach 1 in the thermodynamic limit $L\rightarrow \infty$. }
\label{Figfra}
\end{center}
\end{figure}

To quantify Hilbert-space fragmentation, it is useful to examine the Hamiltonian directly in the many-body Fock basis, where fragmentation manifests as a block-diagonal structure. A convenient and parameter-independent way to characterize this structure is to recast the problem in terms of graph theory. We associate each Fock basis state $|b\rangle=|\{n_i\}\rangle$ ($n_i$ is the boson occupation on site $i=-k,\cdots,0,\cdots,k$) with a node of a graph, and we draw an (undirected) edge between two nodes $|b\rangle$ and $|b^\prime\rangle$ ($b\neq b^\prime$) whenever the Hamiltonian has a nonzero off-diagonal matrix element, $H_{bb^\prime}\neq 0$ (equivalently, whenever the dynamics allows a transition between $|b\rangle$ and $|b^\prime\rangle$). The connected components of this graph then define the Krylov sectors. This is because those Fock basis states belonging to different components cannot be reached from one another by repeated action of the Hamiltonian, implying that the Hamiltonian matrix decomposes into dynamically disconnected blocks once the basis is ordered by component. In this language, Hilbert-space fragmentation corresponds to the graph splitting into many components, while the sizes of these components quantify whether the fragmentation is strong (many comparable components) or weak (one giant component plus many small ones).

In practice, we construct the graph generated by the Hamiltonian in the many-body Fock basis within a fixed symmetry sector $(N, P)=(2L,0)$ and decompose it into disconnected components (blocks). As shown in the upper inset of Fig.~\ref{Figfra}, the block diagonal structure of the Hamiltonian matrix clearly indicates the Hilbert-space fragmentation. The number of fragmented blocks is shown in the main panel of Fig.\ref{Figfra}, revealing an exponential increase with the lattice size $L$. Besides, the lower inset of Fig.\ref{Figfra} highlights the presence of a dominant block among these fragments, whose dimension closely approaches that of the symmetry sector under consideration. For this reason, although the Hilbert space of the symmetry sector $(N, P)=(2L,0)$ decomposes into an exponentially large number of Krylov sectors in the Fock basis, the dominant block accounts for the overwhelming majority of the Hilbert-space dimension. We thus characterize this phenomenon as weak Hilbert-space fragmentation.

\subsection{Frozen states}
\begin{figure}[htbp]
\begin{center}
\includegraphics[width=0.45\textwidth]{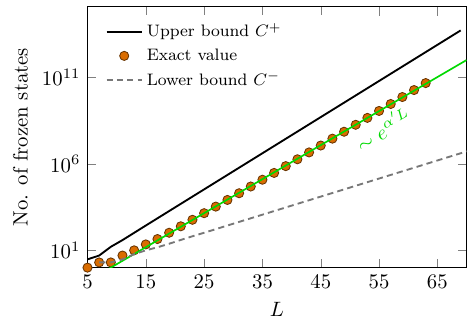}
\caption{(Color online) The number of frozen states corresponds to invariant subspaces of dimension 1 in the symmetry sector $(N, P)=(2L, 0)$ and scales exponentially with the system size $L$ (No. of frozen states $\sim e^{\gamma^\prime L}$, $\gamma^\prime \approx 0.453$ is the fitting parameter). We compare the resulting upper and lower bounds with exact numerical results. Because we only need the count number of frozen states, we use a dynamic programming algorithm to compute it efficiently, thereby avoiding the computational limitations of ED. For $L=7$ and $9$, the numerical results are exactly equal to the lower bound $C^-$.}
\label{fron}
\end{center}
\end{figure}
We are now constructing a family of exponentially many exact eigenstates of the Hamiltonian, which are actually the Fock basis states. These states are known as frozen states \cite{PhysRevX.10.011047}. According to the definition  \cite{PhysRevX.10.011047}, a Fock basis state $|b\rangle=|\{n_i\}\rangle$ can be a frozen state only if 
\begin{equation}
\hat{H}_J|b\rangle=0,
\end{equation}
where $\hat{H}_J$ is the hopping term of the Hamiltonian (\ref{H}). 
The simplest example is the vacuum state $|\mathrm{Vac}\rangle=|0,\cdots,0\rangle$. From the hopping process described by $\hat{b}_i^{\dagger} \hat{b}_{i+1}^2 \hat{b}_{i+2}^{\dagger}+\text { h.c. } $, we can construct frozen states by adding a single three-site block on top of the vacuum state $|\mathrm{Vac}\rangle$. Examples of such blocks include the configurations $|0,\cdots, 0, \boxed{1,0,0}, 0, \cdots,0\rangle$, $|0,\cdots, 0, \boxed{0,0,1}, 0, \cdots,0\rangle$, $|0,\cdots, 0, \boxed{0,1,1}, 0, \cdots,0\rangle$, $|0,\cdots, 0, \boxed{1,1,0}, 0, \cdots,0\rangle$, and $|0,\cdots, 0, \boxed{0,1,0}, 0, \cdots,0\rangle$. Additional frozen states may be formed by adding more non-overlapping blocks on top of $|\mathrm{Vac}\rangle$, while avoiding configurations that allow hopping, such as $|0,\cdots, 0, 1, 0, 1, 0, \cdots,0\rangle$ and $|0,\cdots, 0, 1, 1, 1, 0,\cdots, 0\rangle$. We noticed that the boson occupation numbers on each site can only be 0 or 1. But this is not correct for the edge sites. Since we use open boundary conditions, the occupation numbers on the two boundary sites, $n_{-k}$ and $n_k$, can take arbitrary integer values from $0$ to $N$, while the conditions $n_{-k} n_{-k+2}=0$, and $n_k n_{k-2}=0$ must hold to aviod the hopping process from $-k$, $-k+2$ to $-k+1$ and $k$, $k-2$ to $k-1$. 

Overall,  the frozen states in this open boundary chain should satisfy the following conditions: 
\begin{itemize}
\item At the edge sites, $n_{-k}$ and $n_k$ may be any integer from $0$ to $N$;
\item In the bulk ($i=-k+1,\dots,k-1$), each $n_i$ is either $0$ or $1$;
\item For any two sites separated by a distance of $2$, the occupations satisfy $n_j n_{j+2}=0$.
\end{itemize}
What we are interested in is the total number of frozen states under the two conservation laws, $\sum_i{n_i}=N$ and $\sum_i i{n_i}=0$. 

The exact counting $C$ of this problem is difficult to calculate.  However, we can derive an upper bound $C^+$ by relaxing the feasibility requirement of $n_{\pm k}$, and a lower bound $C^-$ by additionally adding a reflection symmetry(See Appendix \ref{ApA}). The results are 
\begin{equation}
C^+=F_{k-3}F_{k},
\end{equation}
and
\begin{equation}
C^-=\left\{\begin{array}{cl}
F_{(k-2)/2}F_{(k-2)/2-1}, & k \quad \text{is even} \\
F_{(k-1)/2}F_{(k-1)/2-2}, & k \quad \text{is odd}.
\end{array}\right.
\end{equation}
Where $F$ is the Binet formula for the shifted Fibonacci sequence, 
 \begin{equation}
 F_d=(f_0^{d+2}-(-f_0^{-1})^{d+2})/\sqrt{5}.
 \end{equation}
Here $f_0=(1+\sqrt{5})/2$ is the Golden ratio. 
Numerically, we can apply a dynamic programming (DP) algorithm to exactly solve this counting problem and get $C$ for a larger system beyond the computation limits of the ED. Details about this algorithm can be seen in Appendix \ref{DP}. 

The numerical results are presented in Fig.~\ref{fron}.  As it shows, the number of frozen states grows exponentially with the system size $L$, signaling Hilbert-space fragmentation. However, this does not imply that frozen states dominate the fragmentation. Fig.~\ref{Figfra} already shows that a single block contains more than $98\%$ of the total Hilbert-space dimension. This further confirms that the system exhibits weak rather than strong fragmentation.

\section{From weak ETH to ergodicity breaking} \label{ETH}
\subsection{Half-chain entanglement entropy}
\begin{figure*}[htbp]
\begin{center}
\includegraphics[width=0.9\textwidth]{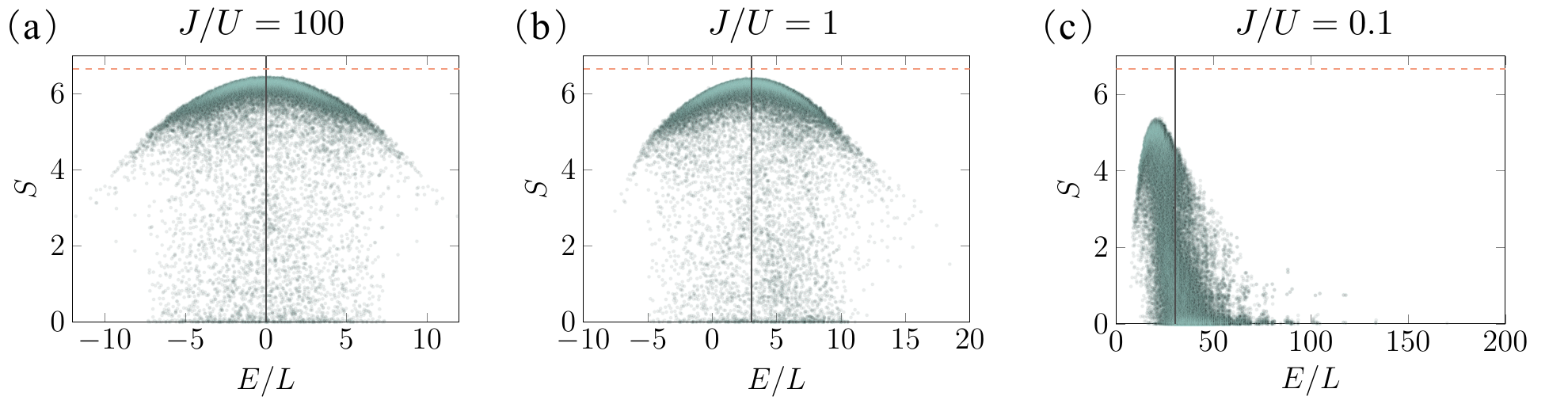}
\caption{(Color online) Half-chain entanglement entropy $S$ vs energy per site $E/L$ for different values of $J/U$ at $L=9$. The horizontal dashed lines are the Page values calculated by averaging 100 random states. The vertical solid lines label the mean energy densities corresponding to each value of $J/U$. (a)-(b) Weak ETH: The entanglement entropy of most eigenstates in the middle energy spectrum is close to, yet slightly below the Page value. Besides, many of the entropies fall on a narrow regime when plotted as a function of energy density, which is a thermalizing feature. While a small number of outlier states, including some in the middle of the spectrum, exhibit much smaller entanglement entropy and are distributed below the Page value. (c) Ergodicity breaking: All the entropies are clearly lower than the page value, and the entropy distribution does not become narrower.  }
\label{Figentropy}
\end{center}
\end{figure*}

As discussed in the previous section, dipole conservation gives rise to weak Hilbert-space fragmentation. This motivates us to examine whether the eigenstates of the system violate the ETH. To this end, we consider the half-chain entanglement entropy, an efficient basis independent tool to identify eigenstate thermalization. If the system obeys ETH, then the entropy is the same for all eigenstates at the same energy density in the thermodynamic limit, which means the entropies should fall on a line when we plot the value as a function of the energy density \cite{PhysRevX.10.011047,PhysRevB.102.144302}. Moreover, the largest entanglement entropies should approach the Page value if ETH is satisfied\cite {PhysRevX.10.011047,PhysRevB.102.144302}. The Page value is obtained by averaging the entanglement entropies over 100 random states within the same symmetry sector $(N, P)=(2L,0)$. Specifically, the random states are chosen as complex vectors $\langle b|\psi_{\mathrm{R}}\rangle = R_1 + iR_2 $, where $(R_1, R_2)$ ​ are independently sampled from the standard normal distribution, and the resulting random states are subsequently normalized.

The half-chain entanglement entropy can be evaluated numerically. We begin by dividing the entire chain into two subsystems, $A$ and $B$. Since the chain length $L$ in our work is always odd, one of them has length $L/2-1$, and the other is $L/2+1$. The half-chain entanglement entropy $S$ for a given eigenstate can then be written as,
\begin{equation}
S_A=-\mathrm{Tr}\rho_A^\alpha\ln\rho_A^\alpha,
\end{equation}
where $\rho_A^\alpha=\mathrm{Tr}_B |\phi_\alpha\rangle \langle \phi_\alpha|$ is the reduced density matrix of subsystem $A$ for the $\alpha$-th eigenstate $|\phi_\alpha\rangle$. Because $S_A$=$S_B$, we thus use $S$ to denote the half-chain entanglement entropy in the following. 

The numerical results of $S$ as functions of energy density $E/L$ are presented in Fig.~\ref{Figentropy}. As one can see in Fig.~\ref{Figentropy}(a) and (b), for a large value of $J/U$, the entanglement entropies of most eigenstates are located in a relatively narrow band. Near the middle of the energy spectrum, these values are close to the Page value, consistent with ergodic behavior and thermalization. These states come from the dominant Krylov sector. At the same time, a small number of outlier states, including some in the middle of the spectrum, exhibit much smaller entanglement entropy and are distributed below the Page value. These non-thermalized states come from the other small Krylov sectors. Therefore, for large values of $J/U$,  the results suggest that the system satisfies the weak ETH, rather than a strong version. It reveals that the weak fragmentation can be compatible with weak ETH. This pheonemenon also have been mentioned in spin-1 system with four-site hopping\cite{PhysRevX.10.011047}. 

While for $J/U\ll 1$, as shown in Fig.~\ref{Figentropy}(c), all the entropies are much smaller than the Page value, and instead fall on a narrow region, they are all distributed below the Page value.  This behavior indicates nonergodic dynamics and suggests that the system does not follow ETH, even in the weak case. This is because the system is close to the atomic limit regime. The eigenstates are approximately occupation number product states rather than becoming chaotic superpositions. It makes the eigenstates only become weakly mixed and leads to low entanglement entropy. This kind of nonergodic phenomenon is also discussed in the traditional Bose-Hubbard model \cite{PhysRevB.102.144302}. 

In addition, the mean energy density is expected to correspond approximately to infinite temperature. Eigenstates near this energy should therefore behave like typical random states in the Hilbert space and have nearly maximal entanglement entropy. This expectation is examined in Figs.~\ref{Figentropy}(a) and \ref{Figentropy}(b), where the eigenstates near the mean energy density show the largest entanglement entropy. It agrees with the ETH picture that the states in the middle of the spectrum are the most thermal and thus have the largest bipartite entanglement.

However, the above simple picture breaks down when the $U$ becomes large. As shown in Fig.~\ref{Figentropy}(c), the maximum of $S$ is shifted away from the mean energy density. This shift occurs because the interaction term dominates the Hamiltonian in the large-$U$ regime and strongly modifies the structure of the density of states (DOS). Instead of forming a single smooth DOS, the eigenstates organize into interaction dominated bands or clusters ( See Appendix.~\ref{DOS}). The hopping term then acts mainly as a weak perturbation and mixes only basis configurations within the same band. Therefore, the center of the full spectrum is no longer necessarily the region where the eigenstates have the largest entanglement.

\begin{figure}[htbp]
\begin{center}
\includegraphics[width=0.45\textwidth]{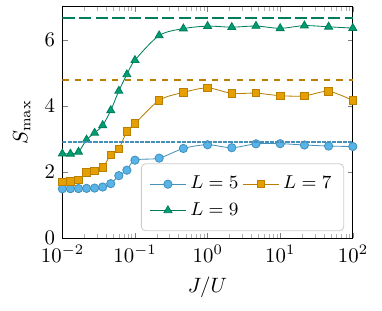}
\caption{(Color online) Maximum half-chain entanglement entropy $S_{\mathrm{max}}$ as a function of $J/U$ for different lattice sizes. The dashed lines indicate the corresponding Page values, with colors matched to each lattice size.}
\label{Smax}
\end{center}
\end{figure}
We further investigate the behavior of the maximum entanglement entropy $S_\mathrm{max}$ as a function of $J/U$. The results are shown in Fig.~\ref{Smax}. For all lattice sizes, $S_{\mathrm{max}}$ increases with $J/U$ from a strongly suppressed value in the interaction dominated regime and then approaches a nearly saturated plateau for $J/U\gtrsim 1$. In this large $J/U$ regime, the plateau value lies close to the corresponding Page value, although a small downward deviation remains, which can be attributed to the influence of fragmented nonthermal sectors. In contrast, when $J/U\ll 1 $, even the largest entanglement entropy remains well below the Page value, indicating that all the eigenstates are far from fully thermal.  These entanglement characteristics support that the system evolves from a nonergodic regime at small $J/U$ to a weak ETH regime at large $J/U$.

\subsection{Autocorrelation function}
\begin{figure}[htbp]
\begin{center}
\includegraphics[width=0.48\textwidth]{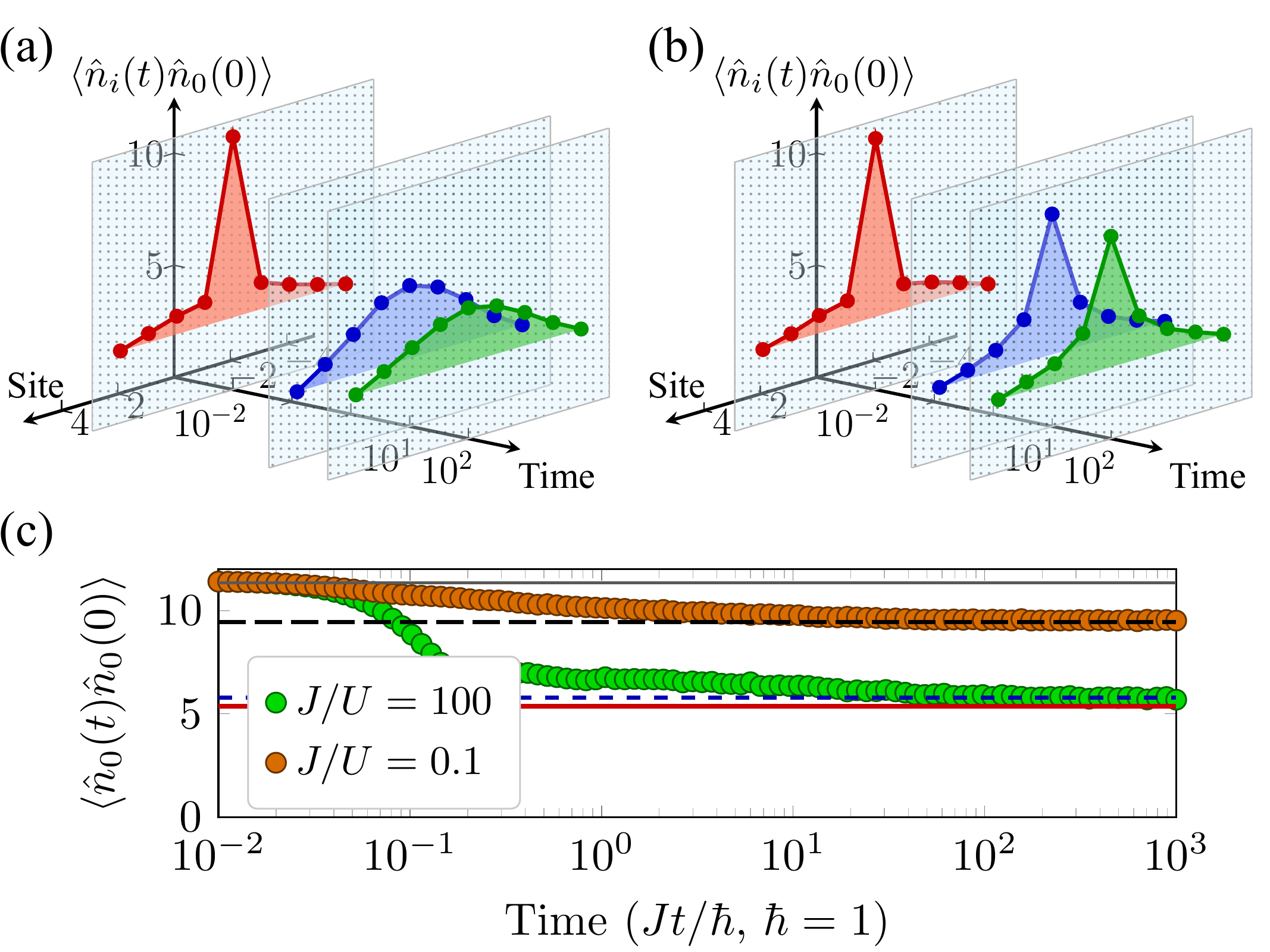}
\caption{(Color online) Time-dependent correlation functions. (a) and (b) show the spatially resolved correlator $\langle \hat{n}_i(t)\hat{n}_0(0)\rangle$ in the symmetry sector $(N, P)=(2L, 0)=(18,0)$ for a random state $|\psi_R\rangle$ corresponding to infinite temperature, with $J/U=100$ and 0.1, respectively. The two cases exhibit qualitatively different long-time dynamics.  (c) shows the autocorrelation function $\langle \hat{n}_0(t)\hat{n}_0(0)\rangle$, illustrating the distinct relaxation dynamics in two regimes. The black solid line indicates the theoretically initial value $C_{\psi_R}(0)$. The long and short dashed lines are the predictions at long-time for $J/U=0.1$ and 100, respectively, which are calculated using Eq.~(\ref{Cinf}). The red solid line is the thermal prediction of the autocorrelator $C_\mathrm{th}$. }
\label{FigTime}
\end{center}
\end{figure}

Another standard quantity that can be used to study the thermalization is the autocorrelation function.  We consider the density correlator $\langle \hat{n}_i(t)\hat{n}_0(0) \rangle$ here, where $i$ labels the lattice site and $\hat{n}_0​(0)$ denotes the local number operator at site 0 and time $t=0$. The expectation value is evaluated for a random state $|\psi_R\rangle$ in the full Hilbert space of the symmetry sector $(N, P)=(2L, 0)$, and the construction of the random state is the same as the previous section. If the ETH holds, the local autocorrelator $\langle \hat{n}_0(t)\hat{n}_0(0)\rangle$ is expected to relax at long times to its thermal value, which should be substantially smaller than its initial value. Otherwise, if the correlator remains anomalously large for long times, the system retains memory of its initial condition and fails to thermalize. 

Numerically, we calculate the real-time evolution of the correlator using the iterative Krylov-space method \cite{TEVO} for a system of size $L=9$. Figs.~\ref{FigTime}(a) and (b) display the spatially resolved correlator at different times. For short times, the correlator is strongly peaked at the central site and decays away from the center in a Lorentzian-like manner. As time evolves, the long-time profile depends sensitively on the ratio $J/U$. For large $J/U$, as Fig.~\ref{FigTime}(a) shows, the sharp central peak gradually broadens, and the correlator approaches a Gaussian-like distribution over the lattice. This behavior indicates the ergodic and ETH features and reveals that the local observables lose memory of their initial conditions and relax toward values determined by thermal equilibrium.  While for small $J/U$, as shown in Fig.~\ref{FigTime}(b), the correlator retains a pronounced peak at the central site even at long times, and the overall profile remains a Lorentzian-like curve. Such persistent spatial inhomogeneity signals the absence of complete thermalization and is therefore indicative of nonergodic behavior. This interpretation is further supported by the autocorrelation function shown in Fig.~\ref{FigTime}(c), which exhibits markedly different long-time behavior in the two regimes. For large $J/U$, the autocorrelation function shows a rapid decay at short times, which means strong thermalization and efficient spread of information. But for small $J/U$,  it decays much more slowly, indicating incomplete thermalization. 

Theoretically, the autocorrrelation function for a random state $|\psi_R\rangle$ at $t=0$ can be writen as,
\begin{equation}
C_{\psi_R}(0)=\langle\psi_R| \hat{n}_0(0) ^2|\psi_R\rangle \approx \frac{1}{D} \operatorname{Tr}_{(N,P)}\left(\hat{n}_0(0) ^2\right),
\end{equation}
which is typically close to the infinite-temperature average in that symmetry sector. Here, $D$ is the Hilbert space dimension of the symmetry sector $(N, P)$. Since the trace is basis independent, we can thus calculate it in the computation Fock basis. For $L=9$, we get 
\begin{equation}
 \frac{1}{D} \operatorname{Tr}_{(N,P)}\left(\hat{n}_0(0) ^2\right) \approx 11.35063.
\end{equation}
Note that the operator $\hat{n}_0(0) ^2$ in the Fock basis is diagonal. This initial value, as shown in Fig.~\ref{FigTime}(c) with a black solid line, is determined only by the operator and the chosen symmetry sector, and therefore does not by itself distinguish between thermal and non-thermal dynamics.

Let us now consider the long-time evolution behavior. We analyze the  long-time average of the autocorrelator,
\begin{equation}
\overline{C_{\psi_R}(\infty)}=\lim _{T \rightarrow \infty} \frac{1}{T} \int_0^T d t\langle\psi_R| \hat{n}_0(t) \hat{n}_0(0)|\psi_R\rangle.
\end{equation}
Since the autocorrelator is calculated for a random state, this is equivalent to the infinite-temperature ensemble average. Thus, for a long time, we have
\begin{align}
\langle\psi_R| \hat{n}_0(t) \hat{n}_0(0)|\psi_R\rangle &\approx \frac{1}{D} \operatorname{Tr}_{(N,P)}\left(\hat{n}_0(t) \hat{n}_0(0)\right) \nonumber\\ 
&=\frac{1}{D} \sum_{\alpha, \beta} e^{i\left(E_\alpha-E_\beta\right) t}\left|\langle \phi_\alpha |\hat{n}_0(0)|\phi_\beta\rangle\right|^2,
\end{align}
where $|\phi_{\alpha,\beta}\rangle$ are the eigenstates of the Hamitonian (\ref{H}), and $\langle \phi_\alpha |\hat{n}_0(0)|\phi_\beta\rangle$ are non-zero because $[\hat{n}_0,\hat{H}]\neq0$. The long-time average of the autocorrelator for a nondegenerate energy spectrum can then be
\begin{equation}\label{Cinf}
\overline{C_{\psi_R}(\infty)} \approx \frac{1}{D} \sum_\alpha\langle\phi_\alpha| \hat{n}_0(0)|\phi_\alpha\rangle^2.
\end{equation}
This value is usually smaller than $C_{\psi_R}(0)$, because at the initial time, the correlator contains all matrix elements of $\hat{n}_0(0)$, while at long times, the oscillating off-diagonal parts cancel out by dephasing. Note that $\overline{C_{\psi_R}(\infty)}$ varies with $J/U$ since it is actually the average of the square of the diagonal matrix element of $\hat{n}_0$ in the energy basis. For $L=9$ and $J/U=100$ (0.1), we obtain $\overline{C_{\psi_R}(\infty)}\approx 5.78071$ (9.45063), which has been show in Fig.~\ref{FigTime}(c) with a short (long) dashed line.  

To assess whether the system thermalizes, we consider the variance of the diagonal matrix elements  of $\hat{n}_0(0)$ in the energy basis,
\begin{align}
\delta &= \overline{C_{\psi_R}(\infty)}  - (\langle \hat{n}_0(0) \rangle_{\mathrm{th}})^2 \nonumber \\ 
& = \frac{1}{D} \sum_\alpha\langle \phi_\alpha| \hat{n}_0(0)|\phi_\alpha\rangle^2-\left(\frac{1}{D} \sum_\alpha\langle \phi_\alpha| \hat{n}_0(0)| \phi_\alpha\rangle\right)^2 \nonumber\\
& = \frac{1}{D} \sum_\alpha\langle \phi_\alpha| \hat{n}_0(0)|\phi_\alpha\rangle^2-\left(\frac{1}{D}  \operatorname{Tr}_{(N,P)}(\hat{n}_0(0))\right)^2.
\end{align}
Here, $\delta$ quantifies the spread of eigenstate expectation values of the local density operator $\hat{n}_0$. 
According to ETH, for a local observable like $\hat{n}_0(0)$, the diagonal matrix elements $\langle\phi_\alpha| \hat{n}_0(0)|\phi_\alpha\rangle$ should become approximately equal for all energy eigenstates at the same energy density. In the context of our infinite-temperature ensemble,  this means these diagonal matrix elements should all cluster tightly around the overall thermal average $\langle \hat{n}_0(0) \rangle_{\mathrm{th}}$. For this reason, if ETH is satisfied, one expects $\delta$ should vanish in the thermodynamic limit, $\lim _{L \rightarrow \infty} \delta=0$. It reveals that Eq.~(\ref{Cinf}) is \textit{not} the value of the autocorrelator at long-time if ETH holds. Instead, the expected thermal value of the autocorrelator should be $C_{\mathrm{th}}=(\langle \hat{n}_0(0) \rangle_{\mathrm{th}})^2$. A vanishing $\delta$ thus signals that the long-time autocorrelation approaches $C_{\mathrm{th}}$ and that the system loses memory of the initial local correlations. $C_{\mathrm{th}}$ is independent of the $J/U$ and it is 5.36718 (show in Fig.~\ref{FigTime}(c) with a red solid line) for $L=9$. 

\begin{figure}[htbp]
\begin{center}
\includegraphics[width=0.45\textwidth]{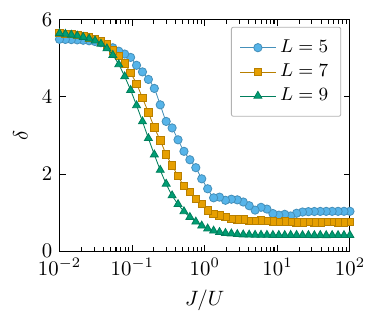}
\caption{(Color online) Difference between the long-time averaged autocorrelator and the thermal value, $\delta$, as a function of $J/U$ for different system sizes. For large $J/U$, $\delta$ decreases toward zero with increasing system size, indicating the ETH behavior in the thermodynamic limit. While for small $J/U$, $\delta$  remains finite, signaling a nonergodic feature.}
\label{Delta}
\end{center}
\end{figure}

To investigate how $\delta$ changes with $J/U$, we numerically evaluate $\delta$ as a function of $J/U$ in Fig.~\ref{Delta} for different lattice sizes. As the figure shows, for large $J/U$, $\delta$ decreases with increasing system size and is expected to vanish in the thermodynamic limit, revealing thermal characteristics. In contrast, for small $J/U$,  $\delta$ remains finite, indicating nonergodic behavior. For all three lattice sizes, $\delta$ decreased as $J/U$ increases, which means the system indeed exists a ergodic-nonergodic evolution when tuning $J/U$. Besides, Fig.~\ref{Delta} suggests a transition, or at least a crossover, between the thermalizing and nonergodic regimes at approximately $J/U\approx 0.1 \sim 0.2$.  However, we are not going to determine this value precisely, since the system sizes accessible by full ED are too small to allow for a reliable finite-size scaling analysis.

\section{Eigenvalue statistics} \label{Eigval}
\begin{figure}[htbp]
\begin{center}
\includegraphics[width=0.45\textwidth]{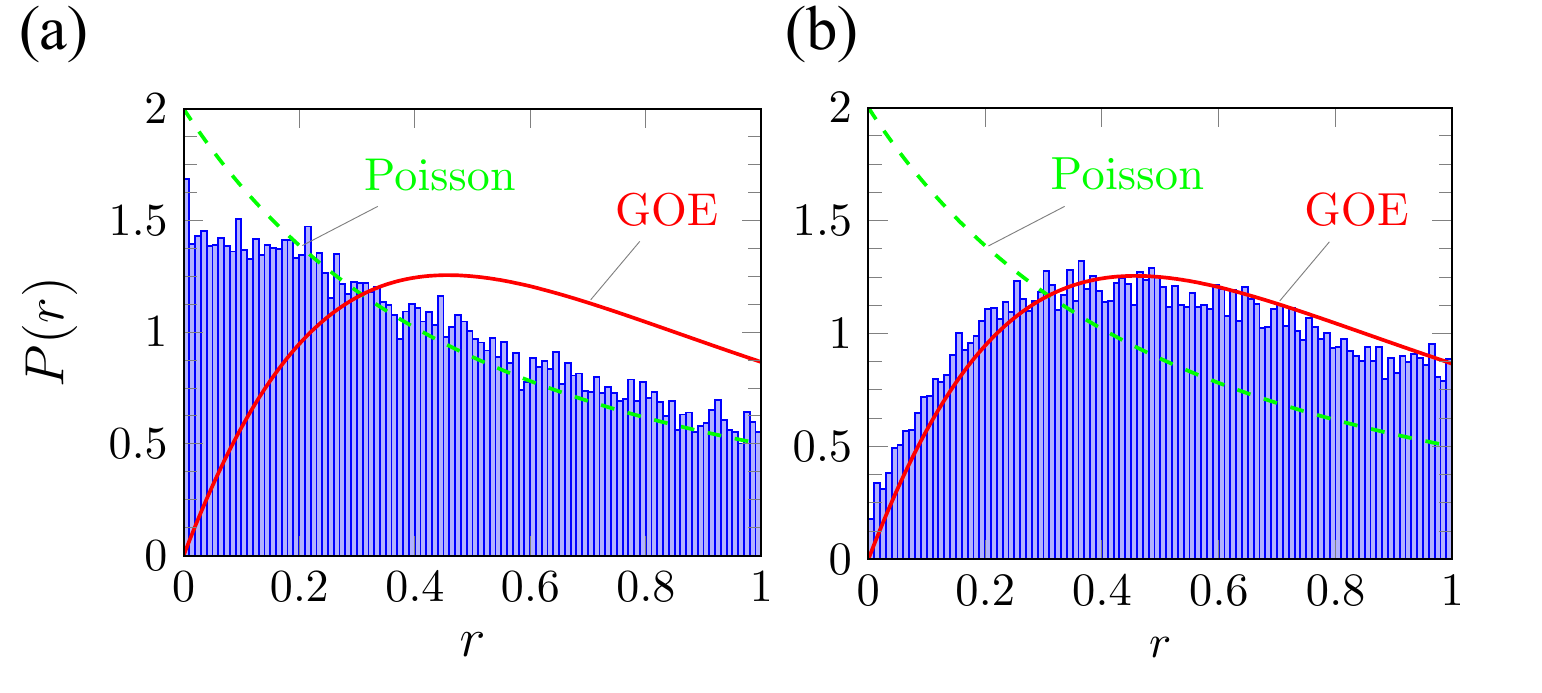}
\caption{(Color online) The distributions of level spacing ratios $r$ with chain length $L=9$ and particle number $N=18$ with dipole moment $P=0$. The distributions are calculated for the entire energy spectrum at interaction strengths (a) $J/U=0.1$ and (b) $J/U=1$. The green dashed line and red solid line represent the theoretical predictions for Poisson statistics (integrable limit) and the Gaussian Orthogonal Ensemble (GOE, chaotic limit), respectively. Here, we partition the $r$ range $[0,1]$ into 100 bins to plot the histogram and normalize it to the probability density function.  }
\label{Fig2}
\end{center}
\end{figure}

\begin{figure*}[htbp]
\begin{center}
\includegraphics[width=0.9\textwidth]{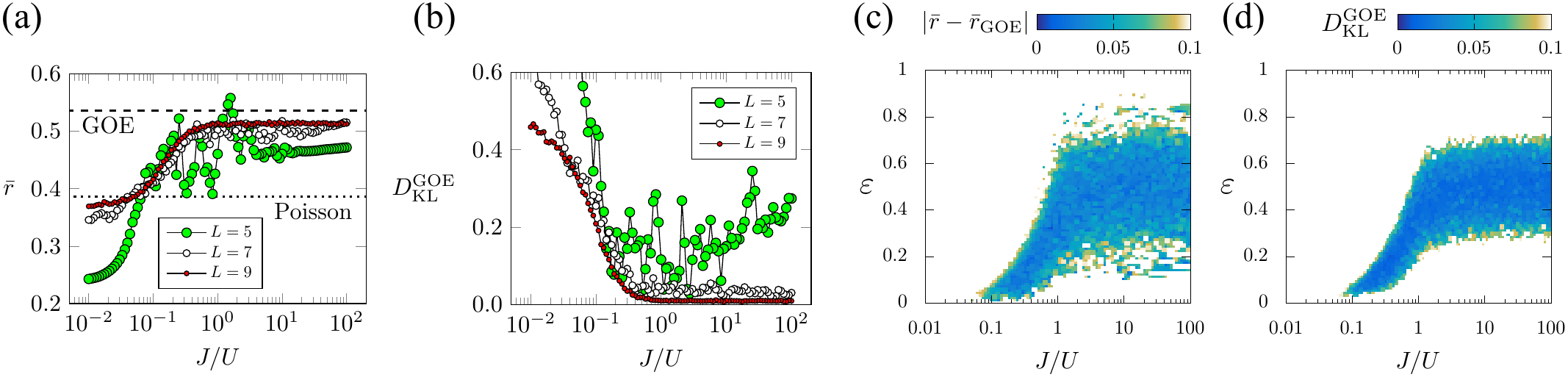}
\caption{(Color online) Level spacing ratio statistics. (a) The mean consecutive level spacing ratio $\bar{r}$ and (b) Kullback-Leibler divergence $D_{\text{KL}}^{\text{GOE}}$ as a function of $J/U$ for lattice sizes $L=5$, 7, and 9, which measuring the deviation between the observed $r$ distribution of the entire energy spectrum and the GOE distribution. The horizontal dashed lines in (a) indicate the theoretical limits for Poisson $\bar{r}_{\text{Poi}}\approx 0.38629$ and GOE $\bar{r}_{\text{GOE}}\approx 0.53590$ statistics. (c)-(d) Level statistics for $L=9$, plotted as a function of $J/U$ and relative energy $\varepsilon=(E-E_{\text{min}})/(E_{\text{max}}-E_{\text{min}})$. (c) The energy-resolved absolute deviation of the mean ratio from the GOE limit, $|\bar{r}-\bar{r}_{\text{GOE}}|$, and (d) the energy-resolved $D_{\text{KL}}^{\text{GOE}}$. The figures (c) and (d) were obtained for 100 equally spaced values of $\log_{10}(J/U) \in [-2,2]$, and divided into 100 windows of equal width $\Delta=0.01$ along the $\varepsilon$ axis. In panels (c) and (d), the color scales are capped at 0.1 to highlight the transition.}
\label{Fig3}
\end{center}
\end{figure*}
In previous sections, we have shown results of several quantities that the system can exhibit in a thermal regime, even though there exists weak Hilbert-space fragmentation. As an important and independent evidence,  in this section, we study the distribution of eigenvalue spacing ratios. We compare this distribution to those characteristic of the Gaussian Orthogonal Ensemble (GOE) and the integrable Poisson statistics, focusing on both the mean value and the full probability distribution. Additionally, we provide details on the numerical methods employed to obtain these results.

\subsection{Definations}
Let $\{E_\alpha\}$ denote the sorted eigenvalues of the Hamiltonian in ascending order, such that $E_\alpha < E_{\alpha+1}$. The consecutive level spacing is defined as $s_\alpha = E_{\alpha+1} - E_\alpha$​. The adjacent spacing ratio is then given by \cite{PhysRevLett.110.084101,PhysRevB.75.155111}:
\begin{equation}
r_\alpha = \min\{\frac{s_\alpha}{s_{\alpha+1}}, \frac{s_{\alpha+1}}{s_\alpha}  \}.
\end{equation}
A primary advantage of analyzing the level spacing ratio statistics is that they are independent of the local density of states. This eliminates the ambiguity associated with the spectral unfolding procedure required for standard level spacing distributions. Furthermore, the ratio is strictly bounded within the interval $r_\alpha \in [0,1]$.

In the context of quantum chaos and thermalizing, the distribution of $r_\alpha$​, denoted as $P(r)$, serves as a robust indicator of the system's dynamical regime. For integrable systems, the energy levels are typically uncorrelated, and the level spacing ratios follow Poissonian statistics. The corresponding probability density function for the ratio is given by:
\begin{equation}
P_{\text{Poi}}(r) = \frac{2}{(1+r)^2},
\end{equation}
which yields a mean value of $\bar{r}_{\text{Poi}} = 2\ln 2 - 1 \approx 0.38629$. Conversely, for chaotic systems exhibiting spectral rigidity and level repulsion, $P(r=0)=0$, the statistics are described by Random Matrix Theory (RMT). Since the Hamiltonian in Eq. (\ref{H}) is real and symmetric (time-reversal invariant), it belongs to the GOE. The distribution of $r$ in this limit is well-approximated by the surrogate expression:
\begin{equation}\label{GOE}
P_{\text{GOE}}(r) = \frac{27}{4} \frac{r+r^2}{(1+r+r^2)^{5/2}},
\end{equation}
with a mean value of $\bar{r}_{\text{GOE}} \approx 0.53590$. The transition from integrability to chaos is thus quantitatively signaled by the evolution of the mean ratio $\bar{r}$ from approximately 0.38629 to 0.53590. 

Furthermore, it is standard practice to analyze ratio distributions over the entire energy spectrum. This allows us to characterize the chaos as a function of energy. To this end, we compute the mean level spacing ratio $\bar{r}$ and the associated distributions within narrow energy windows defined by $[\varepsilon-\Delta,\varepsilon]$. Here, we set the window width $\Delta=0.01$ and define the relative energy as $\varepsilon=(E-E_{\text{min}})/(E_{\text{max}}-E_{\text{min}})$, where $E_{\text{min}}$ and $E_{\text{max}}$ denote the ground state and maximum energy of the spectrum, respectively. Similar energy-resolved analyses of spectral statistics have been successfully employed in previous studies \cite{PhysRevE.103.042209,PhysRevB.91.081103,Castro2021quantumclassical,PhysRevE.107.024210}.

Moreover, to quantify the deviation between the ratio distributions and the GOE prediction, we employ the Kullback-Leibler (KL) divergence \cite{10.1214}. This metric is defined as:
\begin{equation}
D_{\mathrm{KL}}(O \mid R)=\int_{-\infty}^{\infty} O(z) \log \frac{O(z)}{R(z)} \mathrm{d} z, 
\end{equation}
where $O(z)$ is the observed distribution and $R(z)$ serves as the reference distribution.  The divergence vanishes if and only if the distributions are identical. Otherwise, larger values indicate significant deviations between $O$ and $R$. In this work, we compute the divergence relative to the GOE limit, denoted as $D_{\text{KL}}^{\text{GOE}}$ by setting $O$ to the calculated ratio distribution and $R$ to the theoretical GOE distribution.

For the energy-resolved KL divergence, we estimate $D_{\text{KL}}^{\text{GOE}}$ in each energy window $[\varepsilon-\Delta,\varepsilon]$. It is important to note that within each window, the ratio $r$ remains distributed in the range $[0,1]$. Thus, we partition this interval into 20 bins: $[0,0.05],[0.05,0.1],\cdots, [0.95,1]$. Then, we can obtain the distribution of $r$, and after that we compute $D_{\text{KL}}^{\text{GOE}}$ for a given energy window. Throughout this work, when we refer to the mean value $\bar{r}$ and $D_{\text{KL}}^{\text{GOE}}$ at a relative energy $\varepsilon$, we are indicating that these values are calculated in the corresponding energy interval.

\subsection{Numerical results: entire distributions, mean ratio and KL divergence}

In Fig.~\ref{Fig2}, we show the distribution of level spacing ratios for the entire energy spectrum at two values of $J/U$.  For weak hopping amplitude, i.e., $J/U=0.1$, the onsite interaction term dominates the system, and the energy spectrum exhibits nearly uncorrelated level spacings. The distribution is thus close to the Poisson case as shown in Fig.~\ref{Fig2} (a), which reveals an integrable regime. Whereas for strong hopping amplitude, e.g., $J/U=1$, the distribution is close to the GOE distribution as shown in Fig.~\ref{Fig2} (b), which indicates the chaotic and thermal regime, consistent with the previous sections. 

To quantify the change in the level spacing ratio distribution, we consider two other complementary measurements, the mean level spacing ratio $\bar{r}$ and the KL divergence $D_{\text{KL}}^{\text{GOE}}$, which have been shown in Fig.~\ref{Fig3} (a) and (b), respectively. As $J/U$ increases,  $\bar{r}$ evolves from  $\bar{r}_{\text{Poi}}$ to $\bar{r}_{\text{GOE}}$. This becomes clearer for larger system sizes, indicating that the energy spectrum is indeed changing from Poisson-like to GOE-like statistics as the hopping strength increases. 
Besides, $D_{\text{KL}}^{\text{GOE}}$ decreases from a relatively large value at small $J/U$ and approaches zero at large $J/U$, especially for larger lattices. This behavior shows that the full level spacing ratio distribution is close to the GOE prediction at large $J/U$. 

Usually, the level spacing ratio statistics are not uniform across the energy spectrum, and it is therefore useful to perform an energy-resolved analysis. Here, in Fig.~\ref{Fig3}(c) and (d), we plot the quantities $|\bar{r}-\bar{r}_{\text{GOE}}|$ and $D_{\text{KL}}^{\text{GOE}}$ as functions of both $J/U$ and the relative energy $\varepsilon$ for $L=9$ to measuring the difference from the GOE statistics. The results show that the middle part of the energy spectrum changes to GOE more rapidly, while larger deviations persist near the spectrum edges. Such behavior is typical in finite-size systems $J/U$, where edge states are more sensitive to nonuniversal effects and finite-size corrections. Nevertheless, for sufficiently large $J/U$, both the two quantites are small over a broad energy window, confirming that the bulk of the energy spectrum is well described by GOE statistics. All these eigenvalue statistics consistently support the conclusion that, for large $J/U$, the system is chaotic and thermalizing. 

\section{Conclusions}

In this work, we studied a dipole-conserving Bose-Hubbard chain focusing on the symmetry sector $(N, P)=(2L,0)$. We showed that the dipole-conserving constraint leads to Hilbert-space fragmentation. By analyzing the Hamiltonian in the many-body Fock basis, we found that the number of disconnected Krylov sectors grows exponentially with system size. At the same time, the fragmentation is weak rather than strong, because a single dominant block occupies an overwhelming fraction of the Hilbert-space dimension. We also constructed and counted frozen states, whose number increases exponentially with $L$, further confirming the fragmented structure of the constrained Hilbert space.

We then investigated the consequences of this weak fragmentation for eigenstate thermalization and dynamical relaxation. From the half-chain entanglement entropy, we found that for large $J/U$ most eigenstates in the middle of the spectrum have entanglement close to the Page value and form a relatively narrow band as a function of energy density, while only a small number of low-entanglement outliers remain. This behavior is consistent with weak ETH. In contrast, for small $J/U$, the entanglement entropy of all eigenstates is strongly suppressed, indicating nonergodic behavior close to the atomic limit. The real-time density autocorrelation function supports the same picture. For large $J/U$, local density decays efficiently and the difference between the long-time value and the thermal prediction decreases to zero with increasing system size, whereas for small $J/U$, the autocorrelation remains anomalously large at long times, signaling ergodicity breaking.

Finally, the eigenvalue statistics provide independent evidence for the same phenomenon. As $J/U$ increases, the level spacing ratio distribution evolves from Poisson to GOE behavior. Both the mean ratio and the KL divergence show that the full energy spectrum becomes GOE statistics at large $J/U$. The energy-resolved analysis further shows that the middle of the energy spectrum becomes chaotic more rapidly than the spectral edges, as expected in finite systems. 

In summary, these results demonstrate that weak Hilbert-space fragmentation does not by itself prevent chaos or thermalization. Instead, in this bosonic dipole-conserving model, weak fragmentation can coexist with a chaotic and thermalizing regime at large $J/U$, while a nonergodic regime emerges at small $J/U$ due to the interaction dominated atomic limit. Although the accessible system sizes do not allow us to determine whether the change between these two regimes is a sharp transition or a broad crossover, our results establish this model as a simple platform for studying the interplay among kinetic constraints, weak fragmentation, quantum chaos, and ergodicity breaking in interacting many-body systems.

Data and simulation codes are available on Zenodo upon reasonable request \cite{data}.

\section{Acknowledgments}
This research is supported by the Scientific Research Fund of Zhejiang Provincial Education Department under Grant No. Y202248878, the Ph.D. research Startup Foundation of Wenzhou University under Grant No. KZ214001P05, and the open project of the state key laboratory of surface physics in Fudan University (Grant No. KF2022$\_$06).

\appendix
\section{Upper and lower bound of the number of the frozen states}\label{ApA}
As stated in the main text, a Fock state is a frozen state only if it satisfies: 

1. At the edge sites, $n_{-k}$ and $n_k$ may be any integer from $0$ to $N$. In the bulk (i.e., $i=-k+1,\dots,k-1$), the occupation numbers are binary, $n_i\in\{0,1\}$.

2. Distance-2 constraint: For any sites separated by a distance of $2$, the occupations satisfy $n_i n_{i+2}=0$; 

3. The configuration satisfies two global conservation laws, $\sum_i{n_i}=N$ and $\sum_i i{n_i}=0$.

An exact expression for the total number of frozen states is difficult to obtain because the counting problem is not purely local: the admissible configurations $\{n_i\}$ are additionally constrained by two global conservation laws, which couple all sites and substantially modify the combinatorics. Nevertheless, we can derive useful upper and lower bounds.

Let us first consider such a simple counting problem. Given $d$ binary variables,
\begin{equation}
 x_1,x_2,x_3,\cdots, x_d \in \{0,1\},
 \end{equation} 
subject to the nearest-neighbor (distance-1) constraint, 
\begin{equation}\label{Cons2}
 x_i\cdot x_{i+1}=0.
  \end{equation} 
Let $F_d$ denote the number of solutions $\{x_i\}$ satisfying Eq.~(\ref{Cons2}). A recursion for $F_d$ follows by conditioning on the last variable $x_d$. If $x_d=0$, then $x_1,\dots,x_{d-1}$ form any valid length-$d-1$ solution, contributing $F_{d-1}$ possibilities. If $x_d=1$, then necessarily $x_{d-1}=0$, and $x_1,\dots,x_{d-2}$ form any valid length-$d-2$ solution, contributing $F_{d-2}$ possibilities. Since these two cases are disjoint and cover all possibilities, so: 
 \begin{equation}
 F_d=F_{d-1}+F_{d-2}.
 \end{equation}
With initial conditions $F_0=1$ and $F_1=2$ (corresponding to the empty and the length-one binary variable), this sequence is a shifted Fibonacci sequence. For example, $F_2=3$, with solutions $\{x_1=0,x_2=0\}$, $\{x_1=0,x_2=1\}$, and $\{x_1=1,x_2=0\}$. 
 The general expression of $F_d$ is the Binet formula, 
 \begin{equation}
 F_d=(f_0^{d+2}-(-f_0^{-1})^{d+2})/\sqrt{5},
 \end{equation}
 where $f_0=(1+\sqrt{5})/2$ is the Golden ratio. We can conclude that if there are $d$ binary variables satisfying $x_i\cdot x_{i+1}=0$ (distance-1 constraint), then there are $ F_d$ possible solutions. This result will be used below after mapping our constraints onto two independent nearest-neighbor exclusion problems.
 
 For our frozen-state counting problem, the bulk occupations play the role of binary variables, 
 \begin{equation}
 n_{-k+1},n_{-k+2}, \cdots, n_{-1}, n_0, n_1, \cdots, n_{k-2}, n_{k-1}\in \{0,1\}.
 \end{equation}
 Define the bulk particle number and bulk dipole moment as
\begin{equation} \label{SS}
N_{\text{bulk}}=\sum_{i=-k+1}^{k-1}{n_i}, \quad P_{\text{bulk}}=\sum_{i=-k+1}^{k-1}i{n_i}.
\end{equation}
Using the the total particle number $N=2L=2(2k+1)$ and total dipole moment $P=0$, the two edge occupations are determined by the bulk via
\begin{equation}\label{nk}
\left\{\begin{array}{cl}
&n_k+n_{-k}=2(2k+1)-N_{\text{bulk}},\\
&n_k-n_{-k}=-P_{\text{bulk}}/k.
\end{array}\right.
\end{equation}
Thus, once the bulk occupations $n_{-k+1}, \cdots, n_{k-1}$ is fixed, the values of $n_{\pm k}$ are fixed as well. Since $N_{\text{bulk}}\le 2k-1$, we have $n_k+n_{-k}\ge 2k+3>0$, and therefore $n_k$ and $n_{-k}$ both need to greater than 0. This means $n_{-k+2}$ and $n_{k-2}$ must both be zero because of the distance-2 constraint. We thus reduce the problem to counting the number of solutions for bulk binary variables $n_i$ ($i=-k+1,\dots,k-1$) that satisfy
\begin{equation}\label{constaint}
\left\{\begin{array}{cl}
&n_{i}\cdot n_{i+2}=0, \\
&n_{-k+2}=n_{k-2}=0.
\end{array}\right.
\end{equation}
with the additional requirement that the $n_{\pm k}$ determined by Eq.~(\ref{nk}) must be non-zero positive integers.

\textit{Upper bound:}  
To obtain an upper bound, we relax the feasibility requirement on $n_{\pm k}$ and count only bulk binary variables satisfying Eq.~ (\ref{constaint}). The key observation is that the distance-2 constraint factorizes into two independent distance-1 constraints when restricted to the two parity sublattices: even-indexed sites couple only to even-indexed sites, and odd-indexed sites couple only to odd-indexed sites. Equivalently, the constraint $n_i\,n_{i+2}=0$ becomes a distance-1 constraint within each of the two subsequences. For definiteness, when $k$ is odd, the odd- and even-index subsequences of bulk variables may be written as
\begin{equation}\label{odds}
n_{-k+2},n_{-k+4},\dots,n_{-1},n_1,\dots,n_{k-4},n_{k-2},
\end{equation}
and
\begin{equation}\label{evens}
n_{-k+1},n_{-k+3},\dots,n_0,\dots,n_{k-3},n_{k-1},
\end{equation}
respectively. Because $n_{-k+2}=n_{k-2}=0$, the odd-index subsequence effectively contains $k-3$ free binary variables subject to a distance-1 constraint, and hence contributes a factor $F_{k-3}$. The even-index subsequence contains $k$ binary variables subject to the same distance-1 constraint, contributing a factor $F_k$. Since the two subsequences are independent under the distance-2 constraint, the total number of bulk patterns is bounded by the product. The same conclusion holds when $k$ is even. Therefore, an upper bound on the number of frozen states is
\begin{equation}
C^+=F_{k-3}F_{k}.
\end{equation}

\textit{Lower bound:}
For a lower bound, we count a subset of configurations that automatically satisfy the dipole constraint by imposing reflection symmetry,
\begin{equation} \label{re}
n_i=n_{-i}. 
\end{equation}
Under Eq.~(\ref{re}), the bulk dipole moment vanishes identically, $P_{\mathrm{bulk}}=0$, and the bulk particle number reduces to
$N_{\text{bulk}}=n_0+2\sum_{i=1}^{k-1}n_i$. 
Equation~(\ref{nk}) then implies
  \begin{equation}
  n_k=n_{-k}=2k+1-\sum_{i=1}^{k-1}n_i-n_0/2.
   \end{equation}
Since  $n_{\pm k}$ must be integers, $n_0$ must be even, combined with the bulk restriction $n_0\in\{0,1\}$, this forces $n_0$ must be 0. Furthermore, the distance-2 constraint across the center site gives $n_{-1}\cdot n_1=0$ and reflection symmetry implies $n_{-1}=n_1$, hence $n_1=n_{-1}=0$. Altogether, we reduce the lower-bound counting problem to find the number of configurations $\{n_i\}$ ($i=0,1,2,\cdots,k-1$) that satisfy,
 \begin{equation}
\left\{\begin{array}{cl}
&  n_i\cdot n_{i+2}=0,\\
& n_0=n_1=n_{k-2}=0.
\end{array}\right.
\end{equation}
As in the upper-bound analysis, we split the remaining variables into even- and odd-index subsequences, each of which is governed by a distance-1 constraint and thus counted by a Fibonacci number with an appropriate effective length after accounting for the fixed zeros. This yields the lower bound
\begin{equation}
C^-=\left\{\begin{array}{cl}
F_{(k-2)/2}F_{(k-2)/2-1}, & k \quad \text{is even} \\
F_{(k-1)/2}F_{(k-1)/2-2}, & k \quad \text{is odd}
\end{array}\right.
 \end{equation}  
 
 \section{ Dynamic programming algorithm to exactly count the number of frozen states} \label{DP}
 \begin{algorithm}[htbp]
\caption{ Dynamic programming algorithm}
\label{alg:dp-count}
\SetAlgoLined
\KwIn{$L=2k+1$, $N_{\mathrm{target}}$, $P_{\mathrm{target}}$. Sites $i=-k,\ldots,k$.}
\KwOut{$\mathrm{count}$.}

\BlankLine
\textbf{State:} $(u,v)\in\{0,1\}^2$. For each $(u,v)$ maintain a sparse map $M_{uv}[(n,p)]\mapsto $ways\;

\BlankLine
\textbf{Initialize:} $M_{00}[(0,0)]\leftarrow 1$; all other $M_{uv}$ empty\;

\BlankLine
\For{$i\leftarrow -k$ \KwTo $k$}{
    Create empty maps $M'_{uv}$ for all $(u,v)\in\{0,1\}^2$\;
    \ForEach{$(u,v)\in\{0,1\}^2$}{
        \ForEach{$(n,p)\in\mathrm{keys}(M_{uv})$}{
            $\mathrm{ways}\leftarrow M_{uv}[(n,p)]$\;

            \tcp{Choose allowed occupations list $A$}
            \eIf{$u=1$}{
                $A\leftarrow\{0\}$.\;
            }{
                $A\leftarrow \mathrm{site\_allowed}(i,n)$ \tcp*[r]{e.g.\ $\{0,1\}$ or $\{0,\ldots,N_{\mathrm{target}}-n\}$}
            }

            \ForEach{$n_i\in A$}{
                $w\leftarrow (n_i>0)\ ?\ 1:\ 0$\;
                $n^\prime\leftarrow n+n_i$;\quad $p^\prime\leftarrow p^\prime+i n_i$\;
                $r\leftarrow N_{\mathrm{target}}-n^\prime$\;
                \If{$\left|P_{\mathrm{target}}-p^\prime\right|>rk$}{
                    \textbf{continue}\tcp*[r]{optional pruning: skip and jump immediately to the next iteration}
                }
                $(u^\prime,v^\prime)\leftarrow (v,w)$\;
                $M^\prime_{u^\prime v^\prime}[(n^\prime,p^\prime)]\leftarrow M^\prime_{u^\prime v^\prime}[(n^\prime,d^\prime)]+\mathrm{ways}$\;
            }
        }
    }
    $M_{uv}\leftarrow M'_{uv}$ \tcp*[r]{for all $(u,v)$}
}

\BlankLine
$\mathrm{count}\leftarrow \sum_{(u,v)\in\{0,1\}^2} M_{uv}[(N_{\mathrm{target}},D_{\mathrm{target}})]$ \tcp*[r]{missing key treated as 0}
\Return $\mathrm{count}$\;

\end{algorithm}
 We consider integer occupancies $\{n_i\}$ on a one-dimensional lattice of odd length $L=2k+1$ with sites $i \in \{-k,-k+1,\cdots,k\}$. The configuration is constrained by (i) a fixed total particle number, $N=\sum_{i=-k}^kn_i=N_\text{target}$ and a fixed dipole moment, $P=\sum_{i=-k}^kin_i=P_\text{target}$, (ii) bulk sites are hard-core, $n_i\in\{0,1\}$ for $|i|<k$, while boundary sites $n_{\pm k}$ allow arbitrary nonnegative integers subject to the global constraint, (iii)  a distance-2 constraint $n_i\cdot n_{i+2}=0$ ($i+2<k$). 
 
A naive approach to counting the number is to try all possible configurations $\{n_i\}$ and then test whether the global constraints $N=N_\text{target}$ and $P=P_\text{target}$ hold. This becomes impractical because the number of candidate assignments grows exponentially with $L$, and the boundary sites allow many integer values. The runtime would be $\sim e^{L}$, which is called the exponential wall. 

Instead, we use a left-to-right dynamic program that scans sites in increasing order of $i$.  The key observation is that the distance-2 constraint can be enforced locally by remembering only the occupancies of the two previously processed sites. We thus define the binary window state, 
\begin{equation}
(u,v)=(s_{i-2},s_{i-1})\in \{0,1\}^2,
\end{equation}
where $s_i$ is 1 ( $n_i>0$) or 0 ($n_i=0$).  There are a total of 4 possible values of $(u,v)$, and they are: (0,0), (0,1), (1,0), and (1,1). For each window state $(u,v)$, we maintain a table of counts indexed by the partial sum $n=\sum_{x\le i-1} n_x$ and $p=\sum_{x\le i-1}xn_x$. Suppose the number of ways to realize a partial assignment on sites $\{-k,\cdots,i-1\}$ consistent with the constraints is known and denoted by  $W_{i-1}(u,v;n,p)$. Then, the update at site $i$ is:

1. Allowed $n_i$ is determinded by $u$. Let $A(i)$ be the set of allowed values of $n_i$ from the constraints, then if $u=0$, $A(i)=\{0,1\}$ for bulk sites and $A(i)=\{0,1,\cdots, N_\text{target}-n\}$ for boundary sites. While if $u=1$, then only $A(i)=\{0\}$ is permitted.  

2. For each $n_i \in A(i)$, define binary varible $w=s_i$  and update
\begin{equation}
n^\prime = n+n_i, \quad p^\prime = p+in_i,\quad (u^\prime, v^\prime) = (v,w).
\end{equation}
We then accumulate
\begin{equation}
W_{i}(u^\prime, v^\prime; n^\prime, p^\prime)=W_{i}(u^\prime, v^\prime; n^\prime, p^\prime)+W_{i-1}(u, v; n, p).
\end{equation}

3. Feasibility pruning. Let $r=N_\text{target}-n^\prime$ be the number of particles remaining after the update, and note that each future particle can change the dipole moment by at most $k$.  Hence, a necessary condition to still be able to reach $P_\text{target}$ is 
\begin{equation}
|P_\text{target}-p^\prime|\le rk. 
\end{equation}
 A partial configuration violating this bound cannot lead to a valid completion and is discarded. This pruning preserves correctness and substantially reduces the number of tracked states.

4. After processing all sites, the desired count is obtained by summing over the four terminal window states:
\begin{equation}
\text{count}= \sum_{(u,v) \in \{0,1\}^2} W_{k}(u, v; N_\text{target}, P_\text{target}).
\end{equation}
We initialize at the left boundary (before processing any site) with $W_{-k-1}(u=0, v=0; n=0, p=0)=1$ and all other entries zero. 

In practice, $W$ is sparse and only a small subset of $(n,p)$ pairs is reachable. We therefore store each table as an associative map from keys $(n,p)$ to integer counts (“ways”), merging contributions by key during the sweep. The algorithm's runtime is proportional to the number of retained states multiplied by the number of local occupancy choices per site. The pruning criterion above makes this cost manageable for the parameter ranges of interest. For example, if we set $N_\text{target}=2L$ and $P_\text{target}=0$, then the runtime scales as $O(L^4)$.  The above steps have been shown in the Algorithm~\ref {alg:dp-count}.

 \section{The normalized density of states} \label{DOS}

\begin{figure}[htbp]
\begin{center}
\includegraphics[width=0.45\textwidth]{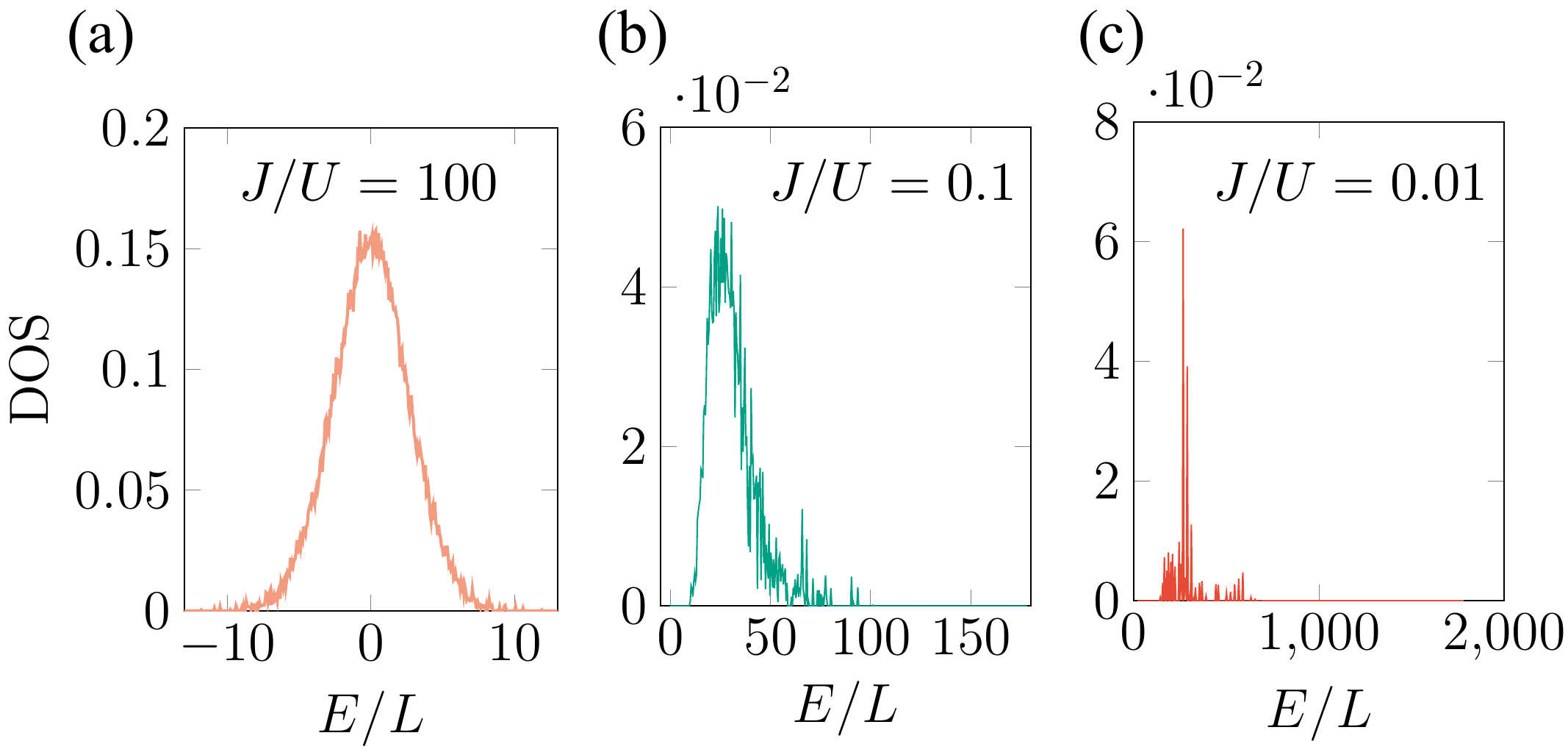}
\caption{(Color online) The normalized Gaussian-broadened density of states for $L=9$. }
\label{FigDOS}
\end{center}
\end{figure}

The density of the states is defined as,
\begin{equation}
\rho({E/L})=\frac{1}{\text{Dim}}\sum_\alpha\delta(E/L-E_\alpha/L).
\end{equation}
Numerically, the $\delta$ function can be broadened using the Gaussian function, 
\begin{equation}
\rho(E/L)=\frac{1}{\text{Dim}}\sum_{\alpha} \frac{1}{\sqrt{2 \pi \sigma^2}} \exp \left(-\frac{\left(E/L-E_{\alpha}/L\right)^2}{2 \sigma^2}\right)
\end{equation}
where $\sigma=0.005$ in our calculations. The results of this Gaussian-broadened density of states (DOS) can be found in Fig.\ref{FigDOS}. In the weakly interacting regime, i.e. $J/U=100$,  the DOS is broad and relatively smooth, consistent with a band-like energy spectrum dominated by hopping processes, even though the system is weak fragmetation. At intermediate coupling, i.e., $J/U=0.1$, the distribution becomes visibly asymmetric and shifts toward positive energy, indicating that interactions substantially restructure the spectral weight. In the strongly interacting regime, i.e., $J/U=100$,   the DOS develops sharp and sparse peaks, signaling the emergence of a more discrete and highly structured spectrum. The figure demonstrates an evolution from a smooth, quasi-continuous energy spectral distribution in the hopping dominated limit to a sharply resolved, interaction dominated spectrum. 

\bibliography{Manuscript}
\end{document}